\documentclass{article}
\usepackage[a4paper]{anysize}
\usepackage{amsthm,amsmath,graphicx}
\usepackage[blocks]{authblk}
  
\usepackage[OT2,T1]{fontenc}
\DeclareSymbolFont{cyrletters}{OT2}{wncyr}{m}{n}
\DeclareMathSymbol{\Sha}{\mathalpha}{cyrletters}{"64}

\usepackage{pgf,tikz}
\usetikzlibrary{arrows}
\usetikzlibrary{intersections}

\newtheorem{thm}{Theorem}

\title{The investigation of singular integro-differential equations
  relating to adhesive contact problems of the theory of
  viscoelasticity}

\author{Nugzar Shavlakadze}

\affil{Tbilisi State University A. Razmadze Mathematical Institute,
  Georgian Technical University, Tamarashvili str. 6, 0177, Tbilisi,
  Georgia e-mail:nusha@rmi.ge, nusha1961@yahoo.com}

\author{Nana Odishelidze}

\affil{Department of Computer Sciences Faculty of Exact and Natural
  Sciences,Iv.Javakhishvili Tbilisi State University, 2, University
  st., 0143,Georgia.  Tfn: 99532304784. E-mail:
  nana\_georgiana@yahoo.com}

\author{Francisco Criado-Aldeanueva\thanks{Corresponding author}}

\affil{Department of Applied Physics II, Polytechnic School, Malaga
  University, Campus Teatinos, s/n (29071), Spain. Tfn:
  +34952132849. E-mail: fcaldeanueva@ctima.uma.es}

\renewcommand{\Re}{\text{Re}\,}
\renewcommand{\Im}{\text{Im}\,}

\newcommand{\ctg}{\text{ctg}}

\begin{document}
\maketitle

\begin{abstract}
  The exact and approximate solutions of singular integro-differential
  equations relating to the problems of interaction of an elastic thin
  finite or infinite non-homogeneous patch with a plate are
  considered, provided that the materials of plate and patch
  possess the creep property. Using the method of orthogonal
  polynomials the problem is reduced to the infinite system of
  Volterra integral equations, and using the method of integral
  transformations this problem is reduced to the different boundary
  value problems of the theory of analytic functions. An asymptotic
  analysis is also performed.
\end{abstract}

The considerable development of the hereditary theory of
Bolzano-Volterra mechanics has been defined by various technical
applications in the theory of metals, plastics and concrete and in
mining engineering. The fundamentals of the theory of viscoelasticity,
the methods for solving linear and non-linear problems of the theory
of creep, the problems of mechanics of inhomogeneously ageing
viscoelastic materials, some bounday value problems of the theory of
growing solids, the contact and mixed problems of the theory of
viscoelasticity for composite inhomogeneosly ageing and
nonlinearly-ageing bodies are considered in \cite{1,2,3,4}.

The full investigation of various possible forms of viscoelastic
relations and of some aspects of the general theory of viscoelasticity are
studied in \cite{5,6,7,8}. Research on the field of creep materials
can be found in \cite{9,10,11,12}.

Contact and mixed boundary value problems on the transfer of the load
from elastic thin-walled elements (stringers, inclusions, patches) to
massive deformable (including aging viscoelastic) bodies, as well as
on the indentation of a rigid stamp into the surface of a viscoelastic
body, represent an urgent problem both in theoretical and applied
aspects. Problems of this type are often encountered in engineering
applications and lead themselves to rigorous mathematical research due
to their applied significance.

\bigskip

Exact and approximate solutions to static contact problems for
different domains, reinforced with non-homogeneous elastic thin
inclusions and patches were obtained, and the behavior of the contact
stresses at the ends of the contact line were investigated in
\cite{13,14,15,16}. One type of analysis assumes continuous
interaction and the other the adhesive contact of thin-shared elements
(stringers or inclusions) with massive deformable bodies. As is known,
stringers and inclusions, such as rigid punches and cuts, are areas of
stress concentration. Therefore, the study of the problems of stress
concentration and the development of various methods for its reduction
is of great importance in engineering practice.

In work \cite{17} we consider integro-differential equations with a
variable coefficient relating to the interaction of an elastic thin
finite inclusion and plate, when the inclusion and plate materials
possess the creep property. Here continuous contact between inclusion
and plate is considered. The solutions to integro-differential
equations of the first order are obtained on the basis of
investigations of different boundary value problems of the theory of
analytic functions. The asymptotic behavior of unknown contact
stresses is established.

In this paper, in contrast to work \cite{17}, contact with a thin
layer of glue is studied when the patch, plate and adhesive materials
have the property of creep. A second order singular
integro-differential equation was obtained. Here the asymptotic
analysis was also carried out and approximate and exact solutions were
obtained for various cases.






\section{Formulation of the problems and  reduction to integral equations}

Let a finite or infinite non-homogeneous patch with modulus of
elasticity $E_1$, thickness $h_1(x)$ and Poisson's coefficient $\nu_1$
be attached to the plate $(E_2, \nu_2)$, which occupies the entire
complex plane and is in the condition of a plane deformation. It is
assumed that the patch, as thin element, is glued to the plate along
the real axis, has no bending rigidity, is in the uniaxial stressed
state and is subject only to tension. The tangential stress $q_0(x)
H(t-t_0)$ acts on the line of contact between the inclusion and the
plate from $t_0$ ($H(t)$ is the unit Heaviside function). The
one-dimensional contact between the plate and patch is affected by a
thin layer of glue with thickness $h_0$ and modulus of shear $G_0$.

It is assumed that the plate, patch and glue layer materials have the
creep property which is characterized by the non-homogeneity of the
ageing process and has different creep measures $C_i(t,\tau) =
\varphi_i(\tau)[1 - e^{-\gamma(t-\tau)}]$, where $\varphi_i(\tau)$ are
the functions that define the ageing process of the plate, patch and
glue layer materials; the age of the different materials is $\tau_i(x)
= \tau_i = \text{const}$; $\gamma=\text{const} > 0$, $i = 1,2,3$.

Besides, the plate Poisson's coefficients for elastic-instant
deformation $\nu_2(t)$ and creep deformation $\nu_2(t,\tau)$ are the
same and constant: $\nu_2(t) = \nu_2(t,\tau) = \nu_2 = \text{const}$.

Assuming that every element of the glue layer is under the condition
of pure shear, the contact condition has the form \cite{18}
\begin{equation}
  u_1(t,x) - u_2(t,x,0) = k_0(I - L_3) q(t,x), \qquad |x| \le 1,
  \label{eq:1.1}
\end{equation}
where $u_2(t,x,y)$ is the displacement of the plate points along the
$ox$-axis and $k_0 := h_0 /G_0$, $u_1(t,x)$ is the displacement of the
inclusion points along the $ox$-axis, $I$ is the unit operator.

We have to define the law of distribution of tangential contact
stresses $q(t,x)$ on the line of contact and the asymptotic behavior
of these stresses at the end of the patch.

To define the unknown contact stresses we obtain the following
integral equation (see \cite{1,2,3,4} )
\begin{equation}
  \begin{aligned}
  \frac{2(1-\nu_2^2)}{\pi E_2}& (I-L_2) \int_{-1}^1 \frac{q(t,y)\,dy}{y-x}\\
  &= \frac 1{E(x)} (I-L_1) \int_{-1}^x [q(t,y)-q_0(y) H(t-t_0)]\,dy-k_0(I-L_3) q'(t,x), \qquad
  |x|<1, \\
  &\int_{-1}^1 [q(t,y)-q_0(y) H(t-t_0)]\,dy = 0
  \end{aligned}
  \label{eq:1.2}
\end{equation}
where  time operators $L_i=1,2,3$ act on an arbitrary function in
the following manner:
\begin{gather*}
  (I-L_i)\psi(t) = \psi(t) - \int_{\tau_i^0}^t K_i(t+\rho_i,\tau+\rho_i) \psi(\tau)\,d\tau, \qquad
  \rho_i= \tau_i - \tau_i^0, \qquad i = 1,2,3, \\
  K_i(t,\tau) = E_i\frac{\partial C_i(t,\tau)}{\partial\tau},  \qquad
  i = 1,2, \qquad
  K_3(t,\tau) = G_0\frac{\partial C_3(t,\tau)}{\partial\tau}, \\
  \omega(t,\tau) = \varphi_3 (\tau)[1 - e^{-\gamma(1-\tau)} ],\qquad
  E(x) = \frac{E_1 h_1(x)}{1 - \nu_1^2},
\end{gather*}
where $\tau_i^0 = t_0$ is the instant of load application.

Introducing the notation
\begin{equation*}
  \varphi(t,x) = \int_{-1}^x[q(t,y)-q_0(y) H(t-t_0)]\,dy, \qquad
  \lambda = \frac{2(1-\nu_2^2)}{E_2}
\end{equation*}
from (\ref{eq:1.2}) we obtain the following two-dimensional
integro-differential equation
\begin{multline}
  \frac \lambda\pi (I-L_2) \int_{-1}^1 \frac{\varphi'(t,y)\,dy}{y-x}
  = \frac 1{E(x)} (I-L_1) \varphi(t,x) - k_0(I-L_3) \varphi''(t,x)+g(t,x), \qquad
  |x| < 1, \\
  g(t,x) = -\frac\lambda\pi (1 - E_2 \varphi_2(t)(1 - e^{-\gamma(t-t_0)}))
  \int_{-1}^1 \frac{q_0(y)\,dy}{y-x} - k_0 q_0'(x)(1-G_0\varphi_3(t)(1-e^{-\gamma(t-t_0)}))
  \label{eq:1.3}
\end{multline}
with conditions
\begin{equation}
  \varphi(t,1)=0, \qquad t\ge t_0
  \label{eq:1.4}
\end{equation}

Thus, the above posed boundary contact problem is reduced to the
solution to singular integro-differential equation (SIDE) with
condition (\ref{eq:1.4}). From the symmetry of the problem, we assume,
that $E(x)$ and $q_0(x)$ are even and odd functions, respectively. The
solution of equation (\ref{eq:1.3}) under condition (\ref{eq:1.4})
with respect to variable $x$ can be sought in the class of even
functions.  Moreover, we assume that function $q_0(x)$ is continuous
in Holder's sense (hereinafter, $H$) and is continuous up to the first
order derivative on an interval $[-1,1]$, i.e. $q_0\in C^1([-1,1])$.

\section{The asymptotic investigation}

Under the assumption that
\begin{equation}
  E(x) = (1 -x^2)^\omega b_0(x),
  \label{eq:2.1}
\end{equation}
where $\omega=\text{const}\ge 0$, $b_0(x) = b_0(-x)$, $b_0 \in C ([-1,1])$,
$b_0(x)\ge c_0 = \text{const} > 0$, the solution to problem
(\ref{eq:1.3}), (\ref{eq:1.4}) will be sought in the class of even
function whose derivative with respect to variable $x$ can be
represented as follows:
\begin{equation}
  \varphi'(t,x) = (1-x^2)^\alpha g_0(t,x), \qquad \alpha > - 1 ,
  \label{eq:2.2}
\end{equation}
where $g_0(t,x)= - g_0(t,-x)$, $g_0\in C^1([-1,1])$, $g_0(t,x)\neq 0$,
$x \in [-1,1]$. $\varphi'(t,x)$ represents the unknown tangential
contact stress.

Introducing the notation
\begin{equation*}
  \Phi_0(x,t) =  \int_{-1}^1 \frac{(1-s^2)^\alpha g_0(t,s)}{s-x}\, ds
\end{equation*}
by virtue of the well-known asymptotic formula \cite{28} we have for $-1<\alpha<0$
\begin{gather*}
  \Phi_0(x,t) = \mp\pi\,\ctg \pi\alpha\, g_0(t,\mp1) 2^\alpha (1\pm x)^\alpha+ \Phi_\pm(x,t),\qquad
  x\to \mp1; \\
  \Phi_\mp(x,t) = \Phi^*_\mp(x,t)(1\pm x)^{\alpha_\pm}, \qquad
  \alpha_\pm = \text{const} > \alpha
\end{gather*}
and for $\alpha = 0$
\begin{equation*}
  \Phi_0(x,t) = \mp g_0(t,\mp1) \ln(1\pm x)+ \widetilde \Phi_{\pm}(x,t),\qquad
  x\to \mp1
\end{equation*}
Functions $\Phi^*_\mp(x,t)$ and $\widetilde\Phi_\mp(x,t)$ satisfy
$(H)$'s condition in a neighborhood of the points $x = \mp1$
respectively.

In case $\alpha > 0$ function $\Phi_0(x,t)$ belongs to the $(H)$ class
in a neighborhood of the points $x =\pm 1$.

In addition, we have \cite{22}
\begin{multline*}
  \int_{-1}^x (1-s^2)^\alpha g_0(t,s)\, ds
  = \frac{2^\alpha(1\pm x)^{\alpha+1}}{\alpha+1} g_0(t,\mp1) \, F(\alpha+1,-\alpha,2+\alpha,(1\pm x)/2)+ G_\mp(x,t),\qquad
  x\to \mp1, \\
  \lim_{x\to \mp1} G_\mp(x,t)(1\pm x)^{-(\alpha+1)} = 0
\end{multline*}
where $F(a,b,c,x)$ is a hypergeometric Gaussian function.

The case $- 1 < \alpha < 0$ is not of interest, since negative values
of the indicator $\alpha$ contradict the physical meaning of condition
(\ref{eq:1.1}).

Let $0 \le \alpha \le 1$, then in a neighborhood of the points $x =
-1$ equation (\ref{eq:1.3}) can be written in the following form
\begin{equation}
  \begin{gathered}
    \shoveleft{(I-L_2 ) \Psi(x,t)
      + \frac{2^\alpha (1+ x)^{2+ \varepsilon}(I-L_1) g_0(-1,t)}{2^\omega(\alpha+1)(1+x)^\omega b_0(-1)}
      + (I-L_1) G_-(x,t)(1+x)^{1+\varepsilon-\alpha}}\\
    \shoveright{-k_0  2^\alpha (1+x)^\varepsilon (I-L_3) \widetilde g_0(-1,t) = g(-1,t)(1+x)^{1+\varepsilon-\alpha}}\\
    \Psi(x,t)=\begin{cases}
      \lambda g_0(-1,t)(1+x)^{1+\varepsilon} \ln(1+x) - \frac{\lambda}{\pi}(1+x)^{1+\varepsilon} \widetilde \Phi_-(x,t), &
      \text{for } \alpha = 0\\
      -\frac\lambda\pi(1+x)^{1+\varepsilon-\alpha} \Phi_0(x,t),&
      \text{for } \alpha\neq 0
    \end{cases}
  \end{gathered}
  \label{eq:2.3}
\end{equation}
where $\varepsilon$ is an arbitrarily small positive number. When
passing to limit $x\to -1$, the analysis of the obtained equations
leads to the necessity of satisfying inequality $2+ \varepsilon
>\omega$, i.e. $\omega\le 2$.

In case $\alpha > 1$ from (\ref{eq:2.3}) it follows that $\alpha
=\omega- 1$.

An analogous result is obtained in the neighborhood of the point $x =
1$.

The obtained results can be formulated as follows:
\begin{thm}
  Assuming that (\ref{eq:2.1}) holds, if problem
  (\ref{eq:1.3}),(\ref{eq:1.4}) has the solution in the form
  (\ref{eq:2.2}), then:
  \begin{itemize}
  \item If $\omega> 2$ then $\alpha =\omega- 1$, ($ \alpha > 1$)
  \item If $\omega \le 2$ then $0 \le \alpha \le 1$.
  \end{itemize}
\end{thm}
  
\paragraph{Conclusion.} If the patch rigidity varies by the law
\begin{equation*}
  E(x) = (1 - x^2)^{n+1/2} b_0(x),
\end{equation*}
where $b_0(x) > 0$ for $|x| \le 1$, $b_0(x) = b_0(-x)$, $n\ge 0$ is
integer, then from the above asymptotic analysis, we obtain:
\begin{equation*}
  \alpha=n-\frac 12, \qquad \text{ for }n = 2 , 3,\ldots
\end{equation*}
and $0 < \alpha < 1$ for $n = 0$ or $n = 1$ (the same result is
obtained for $E(x) = b_0(x) > 0$ or $E(x) = \text{const}$, $|x| \le
1$).

\section{An approximate solution to SIDE (\ref{eq:1.3})}

From the relation
\begin{multline*}
  \frac 1\pi \int_{-1}^1 \frac{(1-s)^\alpha (1+s)^\beta P_m^{(\alpha,\beta)}(s)\,ds}{s-x}
  = \ctg \pi\alpha (1-x)^\alpha(1+x)^\beta P_m^{(\alpha,\beta)}(x) -\\
  \frac{2^{\alpha+ \beta} \Gamma(\alpha)\Gamma(\beta+m+1)}{\pi \Gamma(\alpha+\beta+m+1)}
  F (m+1,-\alpha-\beta-m,1-\alpha,(1-x)/2)
\end{multline*}
obtained by Tricomi \cite{19} for orthogonal Jacobi polynomials
$P_m^{(\alpha,\beta)}(x)$ and from the well-known equality (see
\cite{20}).
\begin{equation*}
  m! P_m^{(\alpha,\beta)} (1-2x) = \frac{\Gamma(\alpha+ m+ 1)}{\Gamma(1+\alpha)} F(\alpha+\beta+m+1,-m,1+\alpha,x)
\end{equation*}
we get the following spectral relation for the Hilbert singular operator
\begin{equation}
  \int_{-1}^1 \frac{(1-s^2)^{n-1/2} P_m^{(n-1/2,n-1/2)}(s)\, ds}{s-x}
  =(-1)^n 2^{2n-1} \pi P_{m+2n-1}^{(1/2-n,1/2-n)}(x),
  \label{eq:3.1}
\end{equation}
where $\Gamma(z)$ is the known Gamma function.

\begin{enumerate}
\item  On the basis of the above asymptotic analysis performed in the cases
  \begin{equation*}
    n = 0; n = 1;\qquad E(x) = b_0(x) > 0; \qquad E(x) = \text{const}, \qquad
    |x| \le 1;
  \end{equation*}
  the solution to equation (\ref{eq:1.3}) will be sought in the form
  \begin{equation}
    \varphi'(t,x)=\sqrt{1-x^2} \sum_{k=1}^\infty X_k(t) P_k^{(1/2,1/2)}(x),
    \label{eq:3.2}
  \end{equation}
  where function $X_k(t)$ has to be defined for $k=1,2,\ldots$.
  
  Using relation (\ref{eq:3.1}) and the Rodrigues formula (see
  \cite{21}) for (\ref{eq:3.2}) we obtain
  \begin{equation}
    \begin{aligned}
      \int_{-1}^1 \frac{\sqrt{1-t^2} P_k^{(1/2,1/2)}(t)\,dt}{t-x} &= - 2 \pi P_{k+1}^{(-1/2,-1/2)}(x),\\
      \varphi(t,x) &=-(1-x^2)^{3/2} \sum_{k=1}^\infty \frac{X_k(t)}{2k} P_{k-1}^{(3/2,3/2)}(x),\\
      \varphi''(t,x) &= -2(1-x^2)^{-1/2} \sum_{k=1}^\infty k X_k(t) P_{k+1}^{(-1/2,-1/2)}(x).
    \end{aligned}
    \label{eq:3.3}
  \end{equation}

  Substituting relation (\ref{eq:3.2}), (\ref{eq:3.3}) into equation
  (\ref{eq:1.3}), we have
  \begin{multline}
    - \frac{(1-x^2)^{3/2}}{E_1(x)} (I-L_1) \sum_{r=1}^\infty \frac{X_k(t)}{2 k} P_{k-1}^{(3/2,3/2)}(x)
    - 2 \lambda_0 (I-L_2) \sum_{k=1}^\infty X_k(t) P_{k+1}^{(-1/2,-1/2)}(x) + \\
    2 k_0(1-x^2)^{-1/2} (I-L_3) \sum_{k=1}^\infty k X_k(t) P_{k+1}^{(-1/2,-1/2)}(x) = g(t,x ), \qquad
    |x| \le 1.
    \label{eq:3.4}
  \end{multline}
  Multiplying both parts of equality (\ref{eq:3.4}) by
  $P_{m+1}^{(-1/2,-1/2)}(x)$ and integrating in the interval $(-1,1)$,
  we obtain an infinite system of Volterra's linear integral equations
  \begin{multline}
    k_0 m\left(\frac{\Gamma(m+3/2)}{\Gamma(m+2)}\right)^2 (I-L_3) X_m(t)
    - \sum_{k=1}^\infty R_{mk}^{(2)} (I-L_2) X_k(t)
    - \sum_{k=1}^\infty \frac{R_{mk}^{(1)}}{k} (I-L_1) X_k(t)
    = g_m(t), \\ m=1,2,\ldots
    \label{eq:3.5}
  \end{multline}
  where
  \begin{align*}
    R_{mk}^{(1)} &= \frac 12 \int_{-1}^1 \frac{(1-x^2)^{3/2}}{E(x)} P_{k-1}^{(3/2,3/2)}(x) P_{m+1}^{(-1/2,-1/2)}(x)\,dx,\\
    R_{mk}^{(2)} &= -2\lambda \int_{-1}^1 P_{k+1}^{(-1/2,-1/2)}(x) P_{m+1}^{(-1/2,-1/2)}(x)\,dx\\
    g_m(t) &= \int_{-1}^1 g(t,x) P_{m+1}^{(-1/2,-1/2)}(x)\,dx.
  \end{align*}
  
  Introducing the notation
  \begin{equation*}
    T_m(t) = \omega_m \left[k_0 X_m(t) - \sum_{k = 1}^\infty \frac{R_{mk}^{(1)}}{k \omega_k} X_k(t) - \sum_{k=1}^\infty \frac{R_{mk}^{(2)}}{\omega_k} X_k(t)\right],
  \end{equation*}
  where
  \begin{equation*}
    \omega_m = m \left(\frac{\Gamma(m+3/2)}{\Gamma(m+2)}\right)^2 \to 1, \qquad m\to\infty
  \end{equation*}
  system (\ref{eq:3.5}) will take the form
  \begin{multline}
    T_m(t) - k_0 \int_{t_0}^t K_3(t-\tau) X_k(\tau)\,d\tau
    + \sum_{k=1}^\infty\frac{R_{mk}^{(1)}}{k\omega_k} \int_{t_0}^t K_1 (t-\tau) X_k(\tau)\, d\tau\\
    + \sum_{k=1}^\infty \frac{R_{mk}^{(2)}}{\omega_k} \int_{t_0}^t K_2 (t-\tau) X_k(\tau)\, d\tau = g_m(t),
    \qquad m=1,2,\ldots
    \label{eq:3.6}
  \end{multline}
  
  In condition $G_0\varphi_3(t) = E_1 \varphi_1(t) = E_2 \varphi_2(t)$
  system (\ref{eq:3.6}) reduces to the following ordinary differential
  equation of second order
  \begin{equation}
    \ddot T_m(t)+\gamma(1+ G_0 \varphi_3(t)) \dot T_m(t) = \ddot g_m(t) + \gamma \dot g_m(t),
    \label{eq:3.7}
  \end{equation}
  with initial conditions:
  \begin{equation*}
    T_m(t_0) = 0, \qquad \dot T_m(t_0) = \dot g_m(t_0)
  \end{equation*}
  The solution to this differential equation gives an infinite system
  of linear algebraic equations with respect to $X_m(t)$,
  $m=1,2,\ldots$
  \begin{equation}
    k_0 X_m(t)
    - \sum_{k=1}^\infty \frac{R_{mk}^{(1)}}{k \omega_k} X_k(t)
    - \sum_{k=1}^\infty \frac{R_{mk}^{(2)}}{\omega_k} X_k(t)
    = \frac{T_m(t)}{\omega_m}
    \label{eq:3.8}
  \end{equation}
  where
  \begin{align*}
    T_m(t)
    &= \dot g_m(t_0) \int_{t_0}^t \frac{d\tau}{\alpha(\tau)}
    + \int_{t_0}^t \frac{d\tau}{\alpha(\tau)} \int_{t_0}^\tau [\ddot g_m(s)  + \gamma \dot g_m(s)] \alpha(s)\, ds,\\
    \alpha(t) &= \exp\int_{t_0}^t \gamma(1+ G_0\varphi_3(s))\,ds
  \end{align*}
  Let us investigate system (\ref{eq:3.8}) for regularity in the class
  of bounded sequences using the known relations for the Chebyshev
  first order polynomials and for the Gamma function \cite{5}
  \begin{gather*}
    P_m^{(-1/2,-1/2)}(x) = \frac{\Gamma(m+1/2)}{\sqrt{\pi}\Gamma(m+1)} T_m(x),\qquad
    T_m(\cos(\theta)) = \cos m\theta, \qquad
    \lim_{m\to\infty} m^{b-a}\frac{\Gamma(m+a)}{\Gamma(m+b)} = 1
  \end{gather*}
  we have
  \begin{multline*}
    R_{mk}^{(2)}
    = - \frac{2\lambda\alpha(k)\beta(m)}{\pi\sqrt{(k+1)(m+1)}} \int_0^\pi \cos(k+1) \theta \cos(m+1)\theta \sin\theta\, d\theta\\
    = - \frac{2\lambda\alpha(k)\beta(k)}{\pi\sqrt{(k+1)(m+1)}} \times
    \begin{cases}
      1 - \frac{1}{(2m+3)(2m+1)}, & k=m\\
      -\frac{(-1)^{k+m}+1}{2} \left[
        \frac 1{(k+m+3)(k+m+1)}+ \frac 1{(k-m+1)(k-m-1)}
        \right], & k\neq m,
    \end{cases}\\
    = \begin{cases}
      O(m^{-1}), & k=m, \quad m\to\infty\\
      O(m^{-5/2}), O(k^{-5/2}), & k\neq m, \quad k,m\to\infty,
    \end{cases},
  \end{multline*}
  where $\alpha(k), \beta(m) \to 1$, when $k,m\to\infty$.
  
  By virtue of the Darboux asymptotic formula (see \cite{8}), we
  obtain analogous estimates for
  \begin{equation*}
    R^{(1)}_{mk} = \begin{cases}
      O(m^{-1}), & k = m , \quad m\to  \infty,\\
      O(m^{-5/2}), O(k^{-1/2}),& k\neq m, \quad k,m\to\infty
    \end{cases}
  \end{equation*}
  and the right-hand side $T_m(t)/\omega_m$  of equation (\ref{eq:3.8}) satisfies at least the estimate
  \begin{equation*}
    \frac{T_m(t)}{\omega_m} = O(m^{-1/2}), m\to\infty
  \end{equation*}
  
\item If $n=2$ the solution to equation (\ref{eq:1.3}) will be sought
  in the form
  \begin{equation}
    \varphi'(t,x)= (1-x^2)^{3/2} \sum_{k=1}^\infty Y_k(t) P_k^{(3/2,3/2)}(x),
    \label{eq:3.9}
  \end{equation}
  where numbers $Y_k$ have to be defined for $k=1,2,\ldots$.
  
  Using the relation arising from (\ref{eq:3.1}) and from the
  Rodrigues formula (see \cite{21}) for the orthogonal Jacobi
  polynomials, we get
  \begin{equation}
    \begin{aligned}
      \frac 1{\pi} \int_{-1}^1 \frac{(1-x^2)^{3/2} P_k^{(3/2,3/2)}(t)\,dt}{t-x} &= - 2\pi P_{k+1}^{(-3/2,-3/2)}(x), \\
      \varphi(t,x) &= -(1-x^2)^{5/2} \sum_{k=1}^\infty\frac{Y_k(t)}{2k} P_{k-1}^{(5/2,5/2)}(x),\\
      \varphi''(t,x) &= - 2(1-x^2)^{1/2} \sum_{k=1}^\infty k Y_k(t) P_{k+1}^{(1/2,1/2)}(x).
    \end{aligned}
    \label{eq:3.10}
  \end{equation}

  Similarly as for system (\ref{eq:3.8}), we obtain
  \begin{equation}
    \delta_mY_{m}(t) - \sum_{k=1}^\infty \left(R^{(3)}_{mk} + \frac{R_{mk}^{(4)}}k\right) Y_k(t) = \widetilde T_m(t),  \qquad m=1,2,\ldots
    \label{eq:3.11}
  \end{equation}
  where
  \begin{equation*}
    \begin{aligned}
      R_{mk}^{(3)} &= -2 \lambda \int_{-1}^1 P_{k+1}^{(-3/2,-3/2)}(x) P_{m+1}^{(1/2,1/2)}(x)\, dx,\\
      R_{mk}^{(4)} &= \frac 12 \int_{-1}^1 \frac 1{b_0(x)} P_{k-1}^{(5/2,5/2)}(x) P_{m+1}^{(1/2,1/2)}(x)\, dx,\\
      \widetilde g_m(t) &= \int_{-1}^1 g(t,x) P_{m+1}^{(1/2,1/2)}(x)\,dx\\
      \delta_m &= 4 k_0 m\left(\frac{\Gamma(m+5/2)}{\Gamma(m+3)}\right)^2 \to 1, \qquad m\to\infty,\\
      \widetilde T_m(t) &= \dot{\widetilde g}_m(t_0) \int_{t_0}^t \frac{d\tau}{\alpha(\tau)} + \int_{t_0}^t \frac{d\tau}{\alpha(\tau)} \int_{t_0}^\tau [\ddot{\widetilde g}_m(s) +  \gamma \dot{\widetilde g}_m(s)] \alpha(s)\,ds.
    \end{aligned}
  \end{equation*}
  Using again the Darboux formula, and the known relation for the
  Chebyshev second order polynomials (see \cite{21,22})
  \begin{equation*}
    P_m^{(1/2,1/2)}(x) = \frac{\Gamma(m+3/2)}{\sqrt{\pi}\Gamma(m+2)} U_m(x), \qquad
    U_m(\cos\theta) = \frac{\sin(n+1)\theta}{\sin\theta},
  \end{equation*}
  we obtain the following estimates:
  \begin{equation*}
    \begin{aligned}
      R_{mk}^{(3)} &= \begin{cases}
        O(m^{-1}), & k = m,\quad m\to\infty,\\
        O(m^{-5/2}), O(k^{-5/2}), & k\neq m,\quad k,m\to\infty,
      \end{cases}\\
      R_{mk}^{(4)} &= \begin{cases}
        O(m^{-1}), & k = m,\quad m\to\infty,\\
        O(m^{-1/2}), O(k^{-1/2}), & k\neq m,\quad k,m\to\infty,
      \end{cases},\\
      \widetilde g_m &= O(m^{-1/2}), \qquad m\to\infty.
    \end{aligned}
  \end{equation*}
  Thus, systems (\ref{eq:3.8}) and (\ref{eq:3.11}) are
  quasi-completely regular for any positive values of parameters $k_0$
  and $\lambda$ in the class of bounded sequences.
  
  On the basis of the Hilbert alternatives \cite{23,24}, if the
  determinants of the corresponding finite systems of linear algebraic
  equations are other than zero, then systems (\ref{eq:3.8}) and
  (\ref{eq:3.11}) will have unique solutions in the class of bounded
  sequences. Therefore, by the equivalence of system (\ref{eq:3.8})
  (or (\ref{eq:3.11})) and SIDE (\ref{eq:1.3}) the latter has a unique
  solution.
\end{enumerate}

\section{Exact solution to SIDE (\ref{eq:1.3})}

\begin{figure}[htbp]
  \begin{center}
    \newcommand{\midarrow}{\tikz\draw[-triangle 45] (0,0) -- +(.1,0);}
    \begin{tikzpicture}[scale=1.6]
      \fill[color=white!90!black]
      (-1,-.5) .. controls +(1.5,-.5) .. ++(3,0) .. controls +(1.5,.15) .. ++(3,0) .. controls +(1.5,-.3) ..
      (8,-.5)  .. controls +(-.2,.5) .. ++(0,1) .. controls +(.2,.5) .. ++(0,1) .. controls +(-.1,.5) ..
      (8,2.5)  .. controls +(-1.5,.5) .. ++(-3,0) .. controls +(-1.5,.15) .. ++(-3,0) .. controls +(-1.5,-.3) ..
      (-1,2.5) .. controls +(-.2,-.5) .. ++(0,-1) .. controls +(.2,-.5) .. ++(0,-1) .. controls +(-.1,-.5) .. cycle;
      \draw[->](0,-.5) -- (0,2.5);
      
      \draw (-.5,0) --
      node[pos=.8] {\midarrow} node[pos=1] {\midarrow}
      (8,0) node[anchor=south] {$X$} +(.5,0) node[anchor=west] {Case B};
      \begin{scope}
        \draw (0,0) -- (1,.2) -- (5,.2) -- (5,-.2) -- (1,-.2) -- cycle; 
        \clip (0,0) -- (1,.2) -- (5,.2) -- (5,-.2) -- (1,-.2) -- cycle;
        \foreach \x in {1,2,3,4} {
          \draw (\x,-.2) -- +(.4,.4);
        }
      \end{scope}
      \draw (5.5,0) -- +(-1,.5) node[anchor=south] {$P \delta (x-1) H(t-t_0)$};
      
      \draw(-.5,1.5) --
      node[pos=.12] {\midarrow}
      node[pos=.24] {\midarrow}
      node[pos=.36] {\midarrow}
      node[pos=.48] {\midarrow}
      node[pos=.6] {\midarrow}
      node[pos=.72] {\midarrow}
      node[pos=.84] {\midarrow}
      node[pos=1] {\midarrow}
      (8,1.5) node[anchor=south] {$X$} +(.5,0) node[anchor=west] {Case A};

      \begin{scope}
        \draw (0,1.5) .. controls (2,1.7) .. (7,1.7) -- (7,1.3)  .. controls (2,1.3) ..  cycle;
        \clip (0,1.5) .. controls (2,1.7) .. (7,1.7) -- (7,1.3)  .. controls (2,1.3) ..  cycle;
        \foreach \dx in {1,...,7} {
          \draw (\dx,1.3) -- (\dx+.4,1.7);
        }
      \end{scope}
      \draw (5,1.5) -- +(-1,.5) node[anchor=south] {$\tau_0(t,x) = \tau_0(x) H(t-t_0)$};
      
    \end{tikzpicture}
  \end{center}
  \caption{Graph of cases A (upper) and B (lower)}\label{fig:1}
\end{figure}
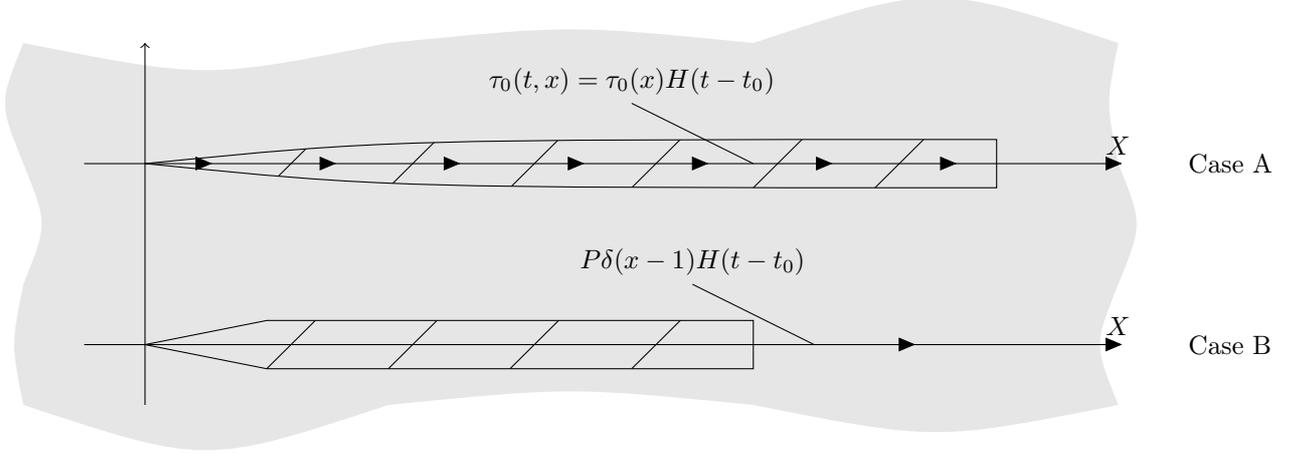

\paragraph{Case A.}
Suppose that a plate on a semi-infinite segment is reinforced by an
inhomogeneous patch whose rigidity changes by the law $E(x) = h
x^2$, $h = \text{const} > 0$. The patch is loaded by a tangential
force of intensity $\tau_0(t,x) = \tau_0(x) H (t-t_0)$ and the plate
is free from external loads (see Figure \ref{fig:1}). We have to
define the law of distribution of tangential contact stresses
$\tau(t,x)$ on the line of contact and the asymptotic behaviour of
these stresses at the end of the patch.
\begin{equation*}
  \tau_0,\tau_0'\in H([0,\infty)), \qquad
    \tau_0 (0) = 0, \qquad
    \tau_0'(x) = O(x^{-2}), \qquad
    x\to  \infty, \qquad
    \int_0^\infty \tau_0(x)\, dx = 0.
\end{equation*}
To determine the unknown contact stresses we obtain the following
integral equation
\begin{equation}
  \begin{gathered}
    (I-L_1)  \frac{\eta_1(t,x)}{h x^2}
    - \frac{\lambda}{\pi} (I-L_2) \int_0^\infty \frac{\eta'_1(t,y)\,dy}{y-x}
    - k_0 (I-L_3) \eta_1''(t,x) = g_1(t,x), \qquad x > 0,\\
    \eta_1(t,0) = 0, \qquad \eta_1(t,\infty) = 0,\\
    \eta_1(t,x) = \int_0^x [\tau(t,y) - \tau_0(t,y)]\, dy, \qquad
    g_1(t,x) = k_0 \tau'_0(t,x) + \frac{\lambda}{\pi} \int_0^\infty\frac{\tau_0(t,y)\,dy}{y-x}\\
    g_1\in H((0,\infty)), \qquad
    g_1(t,x) = O(1), \quad
    x\to 0_+, \qquad
    g_1(t,x) = O(x^{-2}), \quad
    x\to\infty\\
    \eta_1,\eta_1'\in H([0,\infty)), \qquad
      \eta_1''\in H((0,\infty))
  \end{gathered}
  \label{eq:4.1}
\end{equation}

The change of the variables $x = e^\xi$, $y = e^\zeta$ in equation
(\ref{eq:4.1}) gives
\begin{multline}
  (I-L_1) \frac{\varphi_0(t,\xi)}{h e^\xi}
  - \frac\lambda\pi (I-L_2) \int_{-\infty}^\infty \frac{\varphi_0'(t,\zeta)\,d\zeta}{e^{\zeta-\xi}-1}
  - k_0 e^{-\xi}(I-L_3) [\varphi_0''(t,\xi) - \varphi_0'(t,\xi)] \\
  = e^\xi g_0(t,\xi), \qquad |\xi|<\infty
  \label{eq:4.2}
\end{multline}
where $\varphi_0(t,\xi) = \eta_1(t,e^\xi)$, $g_0(t,\xi) =
g_1(t,e^\xi)$, $|g_0(t,\xi)| \le c e^{-|\xi|}$, $|\xi|\to\infty$.

Subjecting both parts of equation (\ref{eq:4.2}) to Fourier's
transformation with respect to $\xi$ \cite{25} and using the
convolution theorem under condition $E_1 \varphi_1 (\tau) =
G_0\varphi_3 (\tau)$, we obtain following boundary condition of the
Carleman-type problem for a strip
\begin{equation}
  (I-L_1) \Phi(t,s+i)
  + \frac{\lambda hs \text{cth} \pi s}{1+k_0 hs (s+i)}(I-L_2) \Phi(t,s)
  = \frac{F(t,s)}{1+k_0 h s (s+i)}, \qquad |s| < \infty
  \label{eq:4.3}
\end{equation}
where
\begin{equation*}
  \Phi(t,s) = \frac {1}{\sqrt{2\pi}} \int_{-\infty}^{\infty} \varphi_0(t,\xi) e^{i \xi s}\,d\xi, \qquad
  F(t,s) = \frac{1}{\sqrt{2\pi}} \int_{-\infty}^\infty e^\xi g_0(t,\xi) e^{i\xi s}\, d\xi
\end{equation*}

Function $F(t,z)$ is holomorphic on strip $-1<\Im z<1$.

The Carleman-type problem for a strip is formulated as follows:

Find a function which is analytic on strip $-1 < \Im z < 1$, (with the
exception of a finite number of points lying on strip $- 1 < \Im z <
0$, at which it has poles), continuously extendable on strip boundary,
vanishes at infinity and satisfies condition (\ref{eq:4.3})
\cite{26,27}.

If we find function $\Phi(t,z)$ which is holomorphic on strip $0 < \Im
z < 1$, extends continuously on the strip boundary and satisfies
condition (\ref{eq:4.3}), then the solution of the problem is the
function
\begin{equation*}
  \Phi_0(t,z) = \begin{cases}
    \Phi(t,z), & 0 \le \Im z < 1\\
    \frac{-\Phi(t,z+i)+F_0(t,z)}{G(x)}, & -1 < \Im z < 0
  \end{cases}
\end{equation*}
where
\begin{equation*}
  G(z) = \frac{\lambda hz \text{cth} \pi z}{1+k_0 hz (z+i)}, \qquad
  F_0(t,z) = \frac{F(t,z)}{1+k_0 hz(z+i)}
\end{equation*}
Representing the function $G(s)$ in the form
\begin{equation*}
  G(s) = \frac{\lambda s}{i k_0(s^2+1)}
  \frac{k_0 h(s^2+1) \text{cth} \pi \text{sth} \frac\pi2 s}{1+ k_0 hs(s+i)}
  \frac{\text{sh}\frac\pi 2 (s+i)}{\text{sh}\frac \pi2 s}
  = \frac{\lambda s}{ik_0 (s^2+1)} G_0(s) \frac{\text{sh}\frac\pi2 (s+i)}{\text{sh}\frac\pi 2 s},
\end{equation*}
where
\begin{equation*}
  G_0(s) = \frac{k_0 h(s^2+1) \text{cth}\pi \text{sth}\frac\pi2 s}{1+ k_0 hs(s+i)}.
\end{equation*}
and remarking that the index of function $G_0(s)$ on
$(-\infty,\infty)$ is equal to zero and $G_0(s)\to 1$,
$s\to\pm\infty$, function $\ln G_0(s)$ is integrable on the axis and
we can write it in the form
\begin{equation}
  G_0(s) = \frac{X_0(s+i)}{X_0(s)}, \qquad |s| <\infty,
  \label{eq:4.4}
\end{equation}
where
\begin{equation*}
  X_0(z) = \exp\left\{
  \frac 1{2i} \int_{-\infty}^\infty \ln G_0(s) \text{cth}\pi (s - z)\, ds \right\}.
\end{equation*}
Function $X_0(z)$ is holomorphic on strip $0 < \Im z < 1$ and bounded
on the closed strip.

Substituting (\ref{eq:4.4}) in condition (\ref{eq:4.3}) and
introducing the notations
\begin{gather*}
  \Psi(t,z) = \frac{z\Phi(t,z)}{X_1(z)},\qquad
  \lambda_0=\frac{k_0}\lambda, \qquad
  X_1(z) = X_0(z) X(z) \text{sh} \frac{\pi z}2, \\
  X(z) = \lambda_0^{iz}\Gamma(2+iz), \qquad
  F(t,z) = \frac{(z+i) F_0(t,z)}{X_1(z+i)},
\end{gather*}
we have
\begin{equation}
  (I-L_1) \Psi(t,s+i)+ (I-L_2) \Psi(t,s) = F(t,s),\qquad |s| < \infty,
  \label{eq:4.5}
\end{equation}
Using Stirling's formula \cite{22} for the Gamma-function, the
following estimate is valid
\begin{equation*}
  |X(z)| = O (|s|^{3/2-\omega}) e^{-\pi|s|/2}, \qquad
  |X_1(z)| = O(|\delta|^{3/2-\omega}), \quad
  z = s+ i\omega, \quad
  0\le\omega\le 1.
\end{equation*}
Applying the Fourier transformation to (\ref{eq:4.5}), we obtain the
Volterra's integral equation of second kind
\begin{equation*}
  [e^w (I-L_1)+ (I-L_2)] \widehat \Phi_1(t,w) = \widehat F(t,w)
  \label{eq:4.6}
\end{equation*}
where $\hat\Phi_1(t,\omega)$, $\hat F(t,\omega)$ are the Fourier
transformations of functions $\Psi(t,s)$, $F_1(t,s)$ respectively.

Since function $F(t,z)$ is analytic on strip $-1 < \Im z < 1$ and
$F(t,z)\to 0$ uniformly, for $|z|\to \infty$, function $\widehat
F(t,w)$ exponentially vanishes at infinity, i.e.  $|\widehat F(t,w)| <
c \exp(-|w|)$, $|w|\to\infty$.

It is easy to show that integral equation (\ref{eq:4.1}) can
equivalently be reduced to the following differential equation of
second order
\begin{equation}
  \ddot{\widehat \Phi}_1(t,w) + \gamma a(t,w) \dot{\widehat \Phi}_1(t,w) = g(t,w)
  \label{eq:4.7}
\end{equation}
with the initial conditions
\begin{align*}
  \widehat \Phi_1(t_0,w) &= \widehat F_1(t_0,w) (1+e^w)^{-1},\\
  \dot{\widehat  \Phi}_1(t_0,w) &= \left[
    \dot{\widehat F}_1(t_0,w) - \gamma \widehat F_1(t_0,w)(e^w \varphi_1(t_0) + \varphi_2(t_0))
    (1+ e^w )^{-1}
    \right]
  (1+e^w)^{-1}
\end{align*}
where
\begin{align*}
  a(t,w) &= 1+(E_1 e^w \varphi_1(t)+ E_2 \varphi_2(t))(1+e^w)^{-1},\\
  g(t,w) &= g_0(t,w)(1+e^w)^{-1},\\
  g_0(t,w) &= \ddot{\widehat F}_1(t,w)+\gamma \dot{\widehat F}_1(t,w).
\end{align*}
Integrating differential equation (\ref{eq:4.7}) and fulfilling the
initial conditions, for function $\widehat \Phi_1(t,w)$ we obtain the
expression
\begin{equation}
  \widehat \Phi_1(t,w) = \{\widehat F_1(t,w)+ F_1(t,t_0,w)\}(1+e^w)^{-1}
  \label{eq:4.8}
\end{equation}
where
\begin{multline*}
  F_1(t,t_0,w)
  =\gamma \widehat F_1(t_0,w) (e^w \varphi_1(t_0)+ \varphi_2(t_0)) (1+e^w)^{-1}
  \int_{\tau_0}^t\exp(-\gamma b(w,\tau,t_0)\,d\tau
  \\
  \shoveright{-\gamma\int_{\tau_0}^t\exp( -\gamma b(w,\tau,t_0)\,d\tau
  \int_{\tau_0}^\tau (\alpha(q,w)-1) \exp(\gamma b(w,q,t_0) \dot{\widehat F}_1(q,w)\,dq,}\\
  b(w,\tau,t_0) = \int_{\tau_0}^\tau a(p,w)\,dp = (\tau-t_0)+(E_1 e^w \psi_1(\tau,t_0)+ E_2 \psi_2(\tau,t_0))(1+e^w)^{-1},\\
  \psi_1(\tau,t_0) = \int_{t_0}^\tau \varphi_1(p)\,dp,\qquad
  \psi_2(\tau,t_0) = \int_{t_0}^\tau\varphi_2(p)\,dp
\end{multline*}
Function $\widehat \Phi_1(t,w)$ given by (\ref{eq:4.8}) has the same
property as function $\widehat F_1(t,w)$ when $|w|\to\infty$.

By the inverse transformation of equality (\ref{eq:4.8}) and using the generalized Parseval's formula
we obtain
\begin{multline}
  \Phi(t,z) = \frac{X_1(z)}{iz} \sqrt{\frac{2}{\pi}}
  \int_{-\infty}^\infty\frac{F(t,s)(is+1)\,ds}{X_1(s-i) \text{sh}\pi(s-z)}\\
  + \frac{X_1(z)}{iz} \gamma(e^w\varphi_2(\tau_0) + \varphi_1(\tau_0)) \int_{t_0}^t Q_1(\tau,z)\,d\tau
  - \frac{X_1(z)}{iz} \gamma \int_{t_0}^t d\tau \int_{t_0}^\tau Q_2(\tau,q,z)\,dq
  \label{eq:4.9}
\end{multline}
where
\begin{align*}
  Q_1(\tau,z) &= \int_{-\infty}^\infty
  \frac{\exp( -\gamma b(w,\tau,\tau_0) \widehat F_1(\tau_0,w) e^{-iwz}\, dw}{(1+e^w)^2},\\
  Q_2(\tau,q,z) &= \int_{-\infty}^\infty
  \frac{\exp( -\gamma b(w,\tau,\tau_0) (\alpha(q,w)-1)\exp(\gamma b(w,q,\tau_0) \dot{\widehat F}_1(q,w) e^{-iwz}\, dw}
  {1+e^w}
\end{align*}

Thus functions $F(t,z)$, $Q_1(\tau,z)$, $Q_2(\tau,q,z)$ are analytic
on strip $-1 < \Im z < 1$ and vanish uniformly $|\Re z|\to
\infty$. The function defined by (\ref{eq:4.9}) is holomorphic on
strip $-1 < \Im z < 0$, and continuously extendable on the strip
boundary.

If function $F(t,z)$ (or $F_0(t,z)$) exponentially vanishes at
infinity, then it is easy to prove that function $\Phi(t,z)$ has the
same property. The inverse Fourier's transformation gives
\begin{equation}\label{eq:4.23}
  \tau(t,x)
  = \tau_0(t,x)+\eta_1'(t,x)
  = \tau_0(t,x)+\frac{x^{-1}}{\sqrt{2\pi}} \int_{-\infty}^\infty is\Phi(t,s) e^{-is\ln x}\,ds.
\end{equation}
Taking into account Cauchy's formula, we get
\begin{multline*}
  \tau(t,x)
  = \tau_0(t,x)+\frac{i x^{-1}}{\sqrt{2\pi}} \int_{-\infty}^\infty (s+i) \Phi(t,s+i) e^{-i(s+i)\ln x}\,ds\\
  = \tau_0(t,x)+\frac{i}{\sqrt{2\pi}}\int_{-\infty}^\infty (s+i) \Phi(t,s+i) e^{-is\ln x}\,ds.
\end{multline*}
Consequently, for the tangential contact stresses have
\begin{equation}
  \tau(t,x) = \tau_0(t,x) + \begin{cases}
    O(1), & x\to 0_+\\
    O(x^{-1-\delta}), & x\to\infty, \quad \delta > 0
  \end{cases}
  \label{eq:4.10}
\end{equation}
The obtained results can be formulated as

\begin{thm}\label{thm:1.a}
If $E(x) = h_0 x^2$, $x > 0$, $h_0 = \text{const} > 0$,
integro-differential equation (\ref{eq:4.1}) has the solution, which
is represented effectively by (\ref{eq:4.23}) and admits estimate
(\ref{eq:4.10}).
\end{thm}

\paragraph{Conclusion 1.}
Thus, when the rigidity of half infinite patch changes with
parabolic law the tangential contact stresses at the thin end of
inclusion has no singularities, it is bounded.

\paragraph{Case B.}
Suppose that on the finite segment of $OX$ axis, the plate is
reinforced by an inhomogeneous patch whose rigidity changes by the
law $E(x) = h x$, $h = \text{const} > 0$ (for example, a wedge shaped
inclusion). The contact between the plate and the patch is
achieved by a thin glue layer with rigidity $k_0(x) = k_0 x$,
$0 < x < 1$, $k_0 = \text{const} > 0$.

The patch is loaded by a horizontal force $P \delta(x-1) H(t-t_0)$
and the plate is free from external loads (see Figure \ref{fig:1}).

To define the unknown contact stresses we obtain the following integral equation
\begin{equation}
  \begin{gathered}
    (I-L_1)\frac{\eta_2(t,x)}{E(x)}
    -\frac{\lambda}{\pi}(I-L_2 ) \int_0^1\frac{\eta_2'(t,y)\,dy}{y-x}
    -(I-L_3)(k_0(x)\eta_2'(t,x))'= 0, \qquad 0<x<1,\\
    \eta_2(t,0) = 0, \qquad
    \eta_2(t,1) = P, \qquad
    \eta_2(t,x)= \int_0^x \tau(t,y)\,dy,\\
    \eta_2 \in H([0,1)),\qquad
    \eta_2' \in C((0,1)),\qquad
    \sup_{x\in(0,1)}|\eta_2'(x)| <  \infty.
  \end{gathered}
  \label{eq:4.11}
\end{equation}
The change of variables $x = e^\xi$, $y=e^\zeta$ in equation
(\ref{eq:4.11}) gives
\begin{equation}
  \begin{gathered}
    (I-L_1 )\frac{\psi(t,\xi)}h
    + \frac\lambda\pi (I-L_2)\int_{-\infty}^0\frac{\psi'(t,\zeta)\,d\zeta}{1-e^{-(\xi-\zeta)}}
    - k_0 (I-L_3) \psi''(t,\xi) = 0, \qquad
    \xi < 0,\\
    \psi(t,-\infty) = 0, \qquad
    \psi(t,0) = P,\qquad
    \psi(t,\xi) =\eta_2(t,e^\xi).
  \end{gathered}
  \label{eq:4.12}
\end{equation}
Applying Fourier's transformation to both parts of equation
(\ref{eq:4.12}) and using the convolution theorem we obtain the
following boundary condition of the Riemann problem \cite{25}
\begin{equation}
  \Psi^+(t,s)
  = (I-L_1) \Phi^-(t,s)+\lambda\text{hscth} \pi s (I-L_2 ) \Phi^-(t,s)
  + k_0 hs^2 (I-L_3 ) \Phi^-(t,s) + g_{01}(t,s), \qquad
  -\infty < s < \infty,
  \label{eq:4.13}
\end{equation}
where
\begin{gather*}
  \Phi^-(t,s) = \frac 1{\sqrt{2\pi}} \int_{-\infty}^0\psi(t,\zeta) e^{is\zeta}\,d\zeta, \qquad
  g_{01}(t,s) = \frac 1{\sqrt{2\pi}} (P i \lambda h(\text{cth}\pi s)_- + P i k_0 h s - k_0 h \psi'(t,0)),\\
  \Psi^+(t,s) = \frac h\pi \int_0^\infty \psi^+(t,\zeta) e^{is\zeta}\,d\zeta, \qquad
  \psi^+(t,\xi) = \begin{cases}
    0, & \xi < 0\\
    \frac\lambda\pi \int_{-\infty}^0 \frac{\psi'(t,\zeta)\,d\zeta}{1-e^{-(\xi-\zeta)}} - k_0 \psi''(t,\xi), & \xi > 0
  \end{cases}
\end{gather*}
Equation (\ref{eq:4.13}) under condition
\begin{equation*}
  G_0 \varphi_3(t) = E_1\varphi_1(t) = E_2 \varphi_2(t)
\end{equation*}
takes the form
\begin{equation}
  \Psi^+(t,s) = (1+\pi\lambda\text{scth} \pi s+ k_0 hs^2)
      [\ddot \Phi(t,s)+\gamma(1+E_1\varphi_1( t+ \rho_1)) \dot \Phi(t,s)]^-
      + g_{01}(t,s)
      \label{eq:4.14}
\end{equation}

The problem can be formulated as follows: it is required to obtain
function $\Psi^+(z)$, holomorphic in the $\Im z > 0$ half-plane, which
vanishes at infinity, and function $\Phi^-(z)$ holomorphic in the $\Im
z < 1$ half-plane (with the exception of a finite number roots of
function $G_1(z)$), which vanishes at infinity. Both are continuous on
the real axis and satisfy condition (\ref{eq:4.14}) \cite{25}.
Boundary condition (\ref{eq:4.14}) is represented in the form
\begin{equation}
  \begin{gathered}
    \frac{\Psi^+(t,s)}{s+i}
    = \frac{G_1(s)}{1+s^2} [\ddot \Phi(t,s)+\gamma(1+E_1\varphi_1(t+\rho_1)) \dot \Phi(t,s)]^-\cdot(s-i)+\frac{g_{01}(t,s)}{s+i}.\\
    G_1(s) = 1+ \lambda \text{hscth} \pi s+ k_0 hs^2,\\
    G_{01}(s) = (k_0h)^{-1} G_1(s)(1+s^2)^{-1}, \quad
    \Re G_{01}(s) > 0, \quad
    G_{01}(\infty) = G_{01}(-\infty) = 1, \quad
    \text{Ind}G_{01}(s) = 0 .
  \end{gathered}
  \label{eq:4.15}
\end{equation}
Introducing the notation
\begin{equation*}
  [\ddot \Phi(t,s) + \gamma(1+E_1\varphi_1(t+\rho_1)) \dot \Phi(t,s)]^- = K^-(t,s)
  \end{equation*}
the solution of this problem has the form \cite{28}
\begin{equation}
  \begin{gathered}
    K^-(t,z)=\frac{\widetilde X(t,z)}{k_0 h(z-i)}, \qquad
    \Im z \le 0, \qquad
    \Psi^+(t,z) = \widetilde X(t,z)(z+i), \qquad
    \Im z > 0,\\
    K^-(t,z) = (\Psi^+(t,z) - g_{01}(t,z)) G_1^{-1}(z), \qquad
    0 < \Im z < 1 ,
  \end{gathered}\label{eq:4.16}
\end{equation}
where
\begin{equation*}
  \widetilde X(t,z) = X(z) \left\{
  \frac 1{2\pi i} \int_{-\infty}^\infty \frac{g_{01}(t,y)\,dy}{X^+(y)(y+i)(y-z)}
  \right\}, \qquad
  X(z) = \exp\left\{
  \frac 1{2\pi i}\int_{-\infty}^\infty \frac{\ln G_{01}(y)\,dy}{y-z}
  \right\}.
\end{equation*}
we have the following differential equation
\begin{equation}
  \ddot\Phi^-( t , s )+\gamma(1+ E_1 \varphi_1(t+\rho_1)) \dot\Phi^-(t,s) = K^-(t,s)
  \label{eq:4.18}
\end{equation}
with the initial conditions
\begin{equation*}
  \Phi^-(t_0,s) = K^-(t_0,s), \qquad
  \dot\Phi^-(t_0,s) = K^-(t_0,s )\gamma E_1\varphi_1(t_0+\rho_1)
\end{equation*}

Integrating differential equation (\ref{eq:4.18}) and fulfilling the
initial condition, for function $\Psi^-(t,s)$ we obtain the expression
\begin{equation}
  \Phi^-(t,s) = K^-(t,s)(1+T(t))
  \label{eq:4.19}
\end{equation}
where
\begin{gather*}
  T(t)
  = \gamma E_1 \varphi_1(t_0+\rho_1) \int_{t_0}^t \exp(-\gamma b(\tau,t_0)\, d\tau
  + \int_{t_0}^t [\exp(-\gamma b(\tau,t_0) \int_{t_0}^\tau \exp(\gamma b(p,t_0)\,dp]\,d\tau,\\
  b(\tau,t_0) = \int_{t_0}^\tau \alpha(q)\,dq,\qquad
  \alpha(q) = 1+ E_1\varphi_1(q+\rho_1)
\end{gather*}
The boundary value of function $Q^-(t,z) =\frac{P}{2\sqrt{2\pi}}
-iz\Phi^-(t,z)$ is the Fourier transform of function
$\Psi'(t,e^\zeta)$. Therefore, we get
\begin{gather}
  \tau(t,x) = \eta_2'(t,x) = \frac 1{\sqrt{2\pi}x} \int_{-\infty}^\infty Q^-(t,s) e^{-is \ln x}\,ds,
  \label{eq:4.20}\\
  \tau(t,x)= O(1), \qquad x\to  1_- \label{eq:4.21}\\
  \tau(t,x)= O(x^{y_0 - 1}), \qquad
  x\to  0_+ , \qquad y_0 > 1/\sqrt{k_0 h}\label{eq:4.22}
\end{gather}

\paragraph{Remark 2.}
If $k_0 h \le 1$, then $\tau(t,x) = O(1)$, $x\to 0_+$.

\paragraph{Remark 3.}
If $k_0 h = 4$, then $G_1(i/2) = 0$ and $\tau(t,x) = O(x^{-1/2})$,
$x\to 0_+$.

Thus, the following theorem is proven:
\begin{thm}
  Integro-differential equation (\ref{eq:4.11}) has the solution,
  which is represented effectively by formula (\ref{eq:4.20}) and
  admits estimates (\ref{eq:4.21}), (\ref{eq:4.22}).
\end{thm}

\section{Discussion and numerical results}

Asymptotic estimates for the solution to integro-differential equation
(\ref{eq:1.2}) are obtained. A method of reduction for infinite
regular systems of linear algebraic equations is justified. For any
law of variation of the stiffness of the patch, tangential contact
stresses have finite values at the ends of patches.

To obtain numerical results, specific values of the aging functions of
the plane, patch and glue are considered in the form
\begin{align*}
  \varphi_1(t) &= 0.0098 \varphi_3(t)\\
  \varphi_2(t) &= 0.00123 \varphi_3(t)\\
  \varphi_3(t) &= 0.09 \cdot 10^{-10}+\frac{4.82 \cdot 10^{-10}}t
\end{align*}
The numerical values of the remaining parameters of the problem are
taken as follows:
\begin{gather*}
  E_1 = 120\cdot 10^9\text{MPa},\quad
  \nu_1 = 0.5,\quad
  E_2 = 95\cdot 10^9\text{MPa},\quad
  \nu_2 = 0.3, \quad\\
  G_0^{(1)} = 0.117 \cdot 10^9\text{MPa},\quad
  (G_0^{(2)} = 11.7 \cdot 10^9 \text{M}Pa),\quad
  h_0 = 5 \cdot 10^{-4} \text{m}, \quad
  h_1(x) = h_1 = 5 \cdot 10^{-2}\text{m},\quad\\
  \gamma= 0.026\,\text{day}^{-1},\quad
  q_0^{(1)}(x) = 10^5\sqrt{1 - x^2} \,\text{N},\quad
  (q_0^{(2)}(x) = 10^7\sqrt{1 - x^2}\,\text{N}), \quad
  \rho_i=0\quad(i=1,2,3),\quad\\
  t_0 = 45\,\text{days},
  t^{(1)} = 2.5 \cdot 10^3\,\text{days}, (t^{(2)}= 9 \cdot 10^3\text{days})
\end{gather*}

The shortened finite systems of linear equations corresponding to
systems (\ref{eq:3.8}) and (\ref{eq:3.11}), consisting of 10 and 12
equations have been solved. The results of the calculation show that
an increase in the number of equations in the systems led to a change
only in the seventh decimal place in the solutions.

Increasing the shear modulus of the glue causes the increase of the
sought contact stresses, and the increase of the time value is
corresponded by a decrease of the values of these stresses.

For comparison, the following should be noted: in contrast to a number
of works in which a rigid contact between two interacting materials is
considered and where unknown contact stresses have singularities at
the ends of the contact line (i.e. stress concentrations arise), in
this work, the contact between two bodies with viscoelastic (creep)
properties is carried out using a thin layer of glue and, therefore,
the found contact stresses at the ends of the contact line turned out
to be limited (finite).

Obviously, the absence of stress concentration in the deformable body is extremely important from a engineering point of view.


\end{document}